\documentclass[twocolumn,showpacs,floatfix,superscriptaddress,amsmath,amssymb,pra]{revtex4}
\usepackage{mathrsfs}
\usepackage{txfonts}
\usepackage{amssymb}
\usepackage{graphicx}
\usepackage{hyperref}
\usepackage{ulem}
 \usepackage{overpic}
 \usepackage{psfrag}
\usepackage{tabularx}
\usepackage{array}
\usepackage{placeins}
\newcommand{\be}{\begin{equation}}
\newcommand{\ee}{\end{equation}}
\newcommand{\PreserveBackslash}[1]{\let\temp=\\#1\let\\=\temp}

\begin{document}
\title{Finite-temperature fidelity and von Neumann entropy
 in the honeycomb spin lattice with quantum Ising interaction}

\author{Yan-Wei Dai}
\affiliation{College of Materials Science and Engineering, Chongqing University, Chongqing 400044, China}
 \affiliation{Centre for Modern Physics,
Chongqing University, Chongqing 400044,  China}

\author{Qian-Qian Shi}
\affiliation{Centre for Modern Physics, Chongqing University,
Chongqing 400044, China} \affiliation{College of Materials Science
and Engineering, Chongqing University, Chongqing 400044, China}


\author{Sam~Young~Cho}
\altaffiliation{E-mail: sycho@cqu.edu.cn}
 \affiliation{Centre for Modern
Physics, Chongqing University, Chongqing 400044, China}
\affiliation{Department of Physics, Chongqing University, Chongqing
400044, China}

\author{Murray T. Batchelor}
 \affiliation{Centre for Modern
 Physics, Chongqing University, Chongqing 400044, China}
\affiliation{Mathematical Sciences Institute and Department of
 Theoretical Physics, Research School of Physics and Engineering,
 The Australian National University, Canberra ACT 2601, Australia}

\author{Huan-Qiang Zhou }
 \affiliation{Centre for Modern
 Physics, Chongqing University, Chongqing 400044, China}
 \affiliation{Department of Physics, Chongqing University, Chongqing
 400044, China}

\begin{abstract}
 The finite temperature phase diagram is obtained for an
 infinite honeycomb lattice with spin-$1/2$ Ising interaction $J$
 by using thermal-state fidelity and von Neumann entropy
 based on the infinite projected entangled pair state algorithm with ancillas.
 The tensor network representation of the fidelity,
 which is defined as an overlap measurement between two thermal states, is presented
 for thermal states on the honeycomb lattice.
 We show that the fidelity per lattice site and the von Neumann entropy can capture
 the phase transition temperatures for applied magnetic field, consistent with
 the transition temperatures obtained via the transverse magnetizations,
 which indicates that a continuous phase transition occurs in the system.
 In the temperature-magnetic field plane, the phase boundary
 is found to have the functional form $(k_BT_c)^2 + h_c^2/2 = a J^2$
 with a single numerical fitting coefficient $a = 2.298$,
 where $T_c$ and $h_c$ are the critical temperature and field with the Boltzmann constant $k_B$.
 For the quantum state at zero temperature, this phase boundary function gives the critical field estimate
 $h_c = \sqrt{2a} J \simeq 2.1438 J$, consistent with the known value $h_c = 2.13250(4)\, J$ calculated from a Cluster Monte Carlo approach.
 The critical temperature in the absence of magnetic field is estimated as $k_BT_c = \sqrt{a}J \simeq 1.5159\, J$,
 consistent with the exact result $k_BT_c =  1.51865...\, J$.
\end{abstract}

 %

\maketitle


 \section {Introduction}
 Since Landau's spontaneous symmetry breaking theory was developed,
 the Landau-Ginzburg-Wilson theory \cite{Landau} has
 been pivotal to understanding
 phase transitions in quantum many-body systems
 \cite{Sachdev,Chaikin}.
 In the last decade, quantum phase transitions
 have been intensively and extensively investigated
 to provide a deeper
 understanding of quantum critical phenomena
 from the perspective of quantum information  \cite{Amico}.
 Significant progress in understanding measures of quantum entanglement, i.e.,
 purely
 quantum correlations absent in classical systems,
 has been achieved in connection with
 quantum phase transitions.
 Especially for any finite-size one-dimensional spin system,
 it was shown that
 the von Neumann entropy quantifies
 the bipartite entanglement between the two partitions
 of the system,
 with logarithmic scaling behavior with respect to the partitioned-system size,
 and the scaling prefactor proportional to the
 central charge $c$, a
 fundamental quantity in conformal field theory and critical phenomena
  \cite{Vidal,Calabrese,Korepin,Tagliacozzo,Pollmann}.
 Recently, geometric measures quantifying multipartite entanglement
 have been shown to scale inversely with the system size  \cite{Shi,Stephan,Hu,JHLiu}
 where the scaling factor is universally connected to
 the minimum Affleck-Ludwig boundary entropy \cite{Affleck}, i.e., the minimum
 groundstate degeneracy  corresponding to one
 of the boundary conformal field theories compatible with
 the bulk criticality  \cite{Liu}.
 Quantum entanglement has then been used
 as a marker and characteristic property of quantum phase transitions driven by
 quantum fluctuations in one-dimensional quantum many-body systems.

 As another way to characterize quantum phase transitions,
 quantum fidelity, defined as an overlap measurement between quantum states,
 has been introduced from
 the basic notion of quantum mechanics
 based on quantum measurement in quantum information
 \cite{Zanardi,Zhou2,Rams,Liu2,Gu3,Gu2,Xiao,Chen,Mukherjee,Gong}.
 In order to explore quantum phase transitions from the viewpoint of quantum fidelity,
 various quantum fidelity approaches have been suggested,
 such as fidelity per lattice site (FLS) \cite{Zhou2}, reduced fidelity \cite{Liu2},
 fidelity susceptibility \cite{Gu3}, density-functional fidelity \cite{Gu2},
 and operator fidelity \cite{Xiao}.
 Quantum fidelity approaches have been shown to capture
 critical behavior in a range of systems and provide
 an alternative marker of quantum phase transitions
 without knowing any detailed broken symmetry.
 Especially, the groundstate FLS has been demonstrated
 to capture drastic changes of the groundstate wave functions
 in the vicinity of a critical point, even for those which cannot be described in
 the framework of Landau-Ginzburg-Wilson theory, such as a
 Beresinskii-Kosterlitz-Thouless transitions \cite{Wang} and topological quantum phase transitions
 \cite{Wang13}
 in quantum one-dimensional many-body systems.
 Further, quantum fidelity has also manifested the relation between
 degenerate groundstates and spontaneous symmetry breaking \cite{Su13,Dai}.

 Such developments in understanding quantum phase transitions
 could be applied towards understanding finite-temperature phase transitions more deeply
 from the perspectives of entanglement and fidelity.
 It is then natural to ask whether such approaches can be generalized to
 characterize finite-temperature phase transitions.
 As a measure of similarity between two quantum states,
 quantum fidelity defined by the overlap function between them
 can be generalized to a fidelity defined by an overlap function between two thermal density matrices
 in thermodynamic systems at finite temperature.
 As is well-known, at zero temperature,
 groundstates in different phases should be orthogonal
 due to their distinguishability in the thermodynamic limit.
 This fact
 allows the quantum fidelity between quantum many-body states in
 different phases signaling quantum phase transitions from an abrupt change of
 the fidelity when system parameters vary through a phase transition point.
 Similar to the quantum fidelity,
 the thermal fidelity may exhibit a singular behavior for a finite-temperature phase transition.
 Very recently, such a thermal fidelity has been studied in the Kitaev honeycomb model \cite{Orus16}.
 A thermal reduced density matrix can be defined from the thermal density matrix.
 For finite-temperature phase transitions,
 a von Neumann entropy defined by the thermal reduced density matrix at finite temperature
 can exhibit a similar behavior to the von Neumann entropy at zero temperature.
 A few investigations
 have been carried out to use the von Neumann entropy
 for finite-temperature phase transitions \cite{JunpengCao,Chamon,Popkov,YangZhao,Porter}.

 In this paper
 we numerically investigate the finite-temperature phase transition
 for the honeycomb lattice with spin-$1/2$ Ising interactions.
 To describe the honeycomb spin lattice,
 we employ the infinite projected entangled pair state (iPEPS) representation
 \cite{iPEPS}
 with ancillas \cite{Czarnik1,Czarnik4}.
 The ancilla states have been introduced to include finite temperature effects.
 Thermal states can be expressed
 in the Hilbert space enlarged due to the ancilla states.
 In terms of a thermal density matrix given by the thermal states,
 we introduce a thermal fidelity and von Neumann entropy at finite temperature.
 We show that the thermal fidelity and von Neumann entropy can detect
 finite-temperature phase transitions.
 The detected phase transition points in the temperature-magnetic field plane
 are discussed by introducing a phase boundary function
 with a single numerical constant.
 From this,
 the estimated quantum critical point at zero temperature
 and the estimated critical temperature in zero magnetic field
 are shown to be consistent with the Monte Carlo calculation \cite{Henk} and the exact result
 \cite{Baxter,Houtappel}, respectively.

 Our paper is organized as follows.
 In Sec. II, we introduce the honeycomb lattice with Ising interactions.
 A brief explanation is given for the extension of the iPEPS
 to a thermal projected entangled pair states (tPEPS)
 with ancillas~\cite{Czarnik1}
 in the enlarged Hilbert space at finite temperature
 on the honeycomb lattice. This approach allows us to define
 a thermal state of the system
 including finite temperature effects.
 In Sec. III, we outline the numerical procedure for the tensor-network-based
 thermal-fidelity and discuss the singular behavior of thermal-fidelity
 indicating the occurrence of a phase transition.
 The singular behavior of the von Neumann entropy
 at the phase transition temperature
 is discussed in Sec. IV.
 The transition temperatures obtained are shown to be consistent
 with those calculated from the magnetization in Sec. V.
 Section VI is devoted to the discussion of
 the phase boundary
 and the estimates of the quantum critical field
 at zero temperature and critical temperature
 in the absence of the magnetic field.
 A summary and remarks are given in Sec. VI.

\section{Honeycomb lattice with quantum Ising interaction}
 We consider  an infinite honeycomb lattice with spin-$1/2$ Ising exchange interaction
 in the presence of a transverse magnetic field.
 The Hamiltonian defined on the honeycomb lattice can be written as
\begin{equation}
  H = H_{zz} + H_x,
    \label{ham}
\end{equation}
 where the spin exchange interaction $H_{zz}$ and the interaction with the magnetic field
 $H_x$ are respectively given by
\begin{subequations}
\begin{eqnarray}
 H_{zz} &=& - J \sum_{\langle s,s' \rangle}\sigma_{z}^{s}\sigma_{z}^{s'},
 \\
 H_x &=&   -h\sum_{s}\sigma_{x}^{s}
\end{eqnarray}
\end{subequations}
 with the strength of the spin exchange interaction $J(>0)$
 and the transverse magnetic field $h$.
 Here $\sigma_{z}^{s}$ and $\sigma_{x}^{s}$ are the spin-$1/2$ Pauli matrices
 at site $s$.  $\langle s,s'\rangle$ runs over all nearest neighbor pairs
 on the honeycomb lattice.
 At zero temperature $T=0$,
 if the spin exchange interaction $J$ is much bigger than the magnetic field $h$, i.e., $J \gg h$,
 the Hamiltonian can be reduced to
 $H \approx - \sum_{\langle s,s'\rangle}\sigma_{z}^{s}\sigma_{z}^{s'}$
 on the honeycomb lattice.
 The Hamiltonian becomes $H \approx  -\sum_{s}\sigma_{x}^{s}$ for $J \ll h$.
 Then the system can undergo a quantum phase transition
 due to a spontaneous $Z_2$-symmetry breaking, which is characterized
 by a non-zero transverse magnetization $M_z = \left\langle \psi| \sigma_z |\psi\right\rangle$
 with a groundstate wavefunction $|\psi\rangle$ at zero temperature.
 The quantum critical point was estimated as $h_c = 2.13250(4)\, J$
 from the Cluster Monte Carlo approach \cite{Henk}.
 The Ising model on the honeycomb lattice has
 the exact critical temperature \cite{Baxter,Houtappel}
\begin{equation}
 k_BT_c= \frac{2}{\log{\left(2+\sqrt{3}\right)}} J= 1.51865 ... \, J
 \label{exact}
\end{equation}
  in the absence of the transverse magnetic field $h=0$.


\subsection {Projected entangled pair states representation
          at finite temperature}

 To study thermal fidelity,
 one needs to first obtain thermal states on the infinite honeycomb lattice
 with the Hamiltonian $H$, where
 every lattice site is described by $S$ spin states
 ($i=1,\dots,S$).
 We then employ iPEPS representation
 with ancillas.
 By appending each lattice with an ancilla,
 i.e., accompanying $a$ ancilla states ($a=1,\dots,S$),
 iPEPS can be extended to thermal projected entangled pair states (tPEPS)
 including finite temperature effects.
 Thus the Hilbert space is enlarged due to the ancilla states.
 Thermal states $|\Psi(\beta)\rangle$ depending on
 temperature can be defined in the enlarged Hilbert space,
 where $\beta$ is the inverse temperature, i.e., $1/\beta=k_BT$
 with the temperature $T$ and the Boltzmann constant $k_B$.
 Thermal states $|\Psi(\beta)\rangle$
 with ancilla states can be obtained from imaginary time evolution \cite{iTEBD}
 of a pure state in the enlarged Hilbert space spanned by
 states $\prod_{s}|i_{s},a_{s}\rangle$,
 where the product runs over all lattice sites $s$.
 Actually, the pure state can be defined as a state at infinite temperature,
 i.e., $|\Psi(0)\rangle =
 \prod_{s}\left( \sum_{i=1}^{S}\frac{1}{\sqrt{S}}|i_{s},i_{a}\rangle \right)$
 because the density of state becomes
 $\rho(\beta=0) = \prod_{s}\left( \sum_{i=1}^{S}\frac{1}{S}|i_{s}\rangle \langle i_{s}| \right)$
 by defining the density of state at finite temperatures \cite{Czarnik1}
 as
\begin{equation}
  \rho(\beta)
  =\mathrm{Tr}_{ancillas}|\Psi(\beta)\rangle\langle\Psi(\beta)|.
    \label{rho}
\end{equation}
 Also,
 the thermal state $|\Psi(\beta)\rangle$ can be written
 in terms of the pure sate $|\Psi(0)\rangle$
 by defining an evolution operator
 $U(\beta)$, i.e.,
\begin{equation}
 |\Psi(\beta)\rangle = U(\beta)|\Psi(0)\rangle.
\end{equation}
 In fact, the density of state at finite temperature
 can be expressed as $\rho(\beta) \propto e^{-\beta H}$
 and then the imaginary time evolution
 for time $\beta$ with $H/2$
 makes it possible to define
 the imaginary time evolution operator
 as $U(\beta) = e^{-\beta H/2}$ for the thermal states
 $|\Psi(\beta)\rangle$.

    \begin{figure}
 \includegraphics[width =0.32\textwidth]{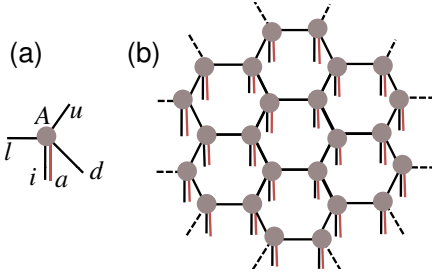}
  \caption{(Color online) (a)
    Pictorial representation of the tensor $A_{lur}^{ia}$.
    (b) Amplitude $\Psi_{A,B}[\{a_{s},i_{s}\}]$ with all bond indices connecting the nearest-neighbors contracted.
    The lines connecting two tensors
    indicate the index contraction.}
\label{fig1}
\end{figure}

 For our honeycomb lattice which is two-site translational invariant,
 a thermal state $|\Psi(\beta)\rangle$ in iPEPS is represented
 by two tensors $A_{lur}^{ia}(\beta)$ and $B_{lrd}^{ia} (\beta)$, where
 spin $S=2$ and
 $l,r,u,d=1,...,D$ are the bond indices
 with the bond dimension $D$.
 In the tensor representation, thermal states can then be
 written as
\begin{equation}
 |\Psi(\beta)\rangle
 =\sum_{a_{s},i_{s}}\Psi_{A,B}[\{a_{s},i_{s}\}]
   \prod_{s}|a_{s},i_{s}\rangle,
   \label{psi}
\end{equation}
 where the sum runs over all indices $i_{s},a_{s}$ at all sites.
 The tensor contraction of the amplitude $\Psi_{A,B}[\{a_{s},i_{s}\}]$
 is shown pictorially on the honeycomb lattice
 in Fig. \ref{fig1}.
 For the imaginary time evolution, the initial state $|\Psi(0)\rangle$
 defined at infinite temperature  $(\beta=0)$
 can be chosen as a product state
 \cite{Czarnik1},
\begin{subequations}
\begin{eqnarray}
  A_{lur}^{ia}(0) &=& \delta^{ia}\delta_{l0}\delta_{u0}\delta_{r0}, \\
  B_{lrd}^{ia}(0) &=& \delta^{ia}\delta_{l0}\delta_{r0}\delta_{d0}
\end{eqnarray}
\end{subequations}
 with the minimal bond dimension $D=1$.
 Thus, once one obtains the tensors $A(\beta)$ and $B(\beta)$
 for a given temperature after the imaginary time evolution,
 the thermal states are determined in the tensor representation.

%

\subsection{Imaginary time evolution and tensor renormalization}
 To calculate a thermal state of the system,
 the idea is to use the imaginary time evolution
 of the initial state $|\Psi(0)\rangle$ at infinite temperature
 driven by the Hamiltonian $H$
 on the honeycomb lattice.
 On performing the imaginary time evolution by
 the time evolution operator $U(\beta) = e^{-\beta H/2}$
 on the initial state $|\Psi(0)\rangle$,
 the second order Suzuki-Trotter decomposition~\cite{Suzuki}
 is employed for an infinitesimal time step as a product
\begin{equation}
  U(d\beta)=U_{x}(d\beta/2)U_{zz}(d\beta)U_{x}(d\beta/2)+O(d\beta^{3}),
    \label{suzuki}
\end{equation}
 where
 the evolution gates of the interaction
 and  of the transverse field
 are defined as $U_{x}(d\beta)=e^{-H_{x}d\beta/2}$ and
 $U_{zz}(d\beta)= e^{-H_{zz}d\beta/2}$, respectively.
 The single-body evolution gate $U_{x}(d\beta/2)$ acting
 on iPEPS with ancillas gives the new tensors $\tilde A$ and $\tilde B$,
\begin{subequations}
   \label{eq8}
\begin{eqnarray}
  \tilde{A}_{lur}^{ia} &\propto& A_{lur}^{ia}+\epsilon \sum_{j=0,1}\sigma_{x}^{ij}A_{lur}^{ja},\\
  \tilde{B}_{lrd}^{ia} &\propto&  B_{lrd}^{ia}+\epsilon \sum_{j=0,1}\sigma_{x}^{ij}B_{lrd}^{ja},
\end{eqnarray}
\end{subequations}
 where $\epsilon=\tanh[hd\beta/4]$ and
 the dimensions of the new tensors
 $\tilde{A}_{lur}^{ia}$ and $\tilde{B}_{lrd}^{ia}$
  are kept as $D$.
 While
 the two-body evolution gate $U_{zz}(d\beta)$ acting on the
 iPEPS with ancillas gives the new tensors $\tilde A$ and $\tilde B$ are
\begin{subequations}

\begin{eqnarray}
  \tilde{A}_{2l+s_{l},2u+s_{u},2r+s_{r}}^{ia}
  &\propto& \epsilon^{s/2} (-1)^{is}A_{lur}^{ia},
     \label{eq9a} \\
  \tilde{B}_{2l+s'_{l},2r+s'_{r},2d+s'_{d}}^{ia}
  &\propto&  \epsilon^{s'/2} (-1)^{is'}B_{lrd}^{ia},
     \label{eq9b}
\end{eqnarray}
\end{subequations}
 where $\epsilon=\tanh[Jd\beta/2]$.
 The indices satisfy $s=s_{l}+s_{u}+s_{r}$ and $s'=s'_{l}+s'_{r}+s'_{d}$
 with $s_{l},s_{u},s_{r},s'_{l},s'_{r},s'_{d}\in\{0,1\}$.
 Equations (\ref{eq9a}) and (\ref{eq9b}) are an exact map
 but
 the tensors $A$ and $B$ are changed from the original $D$-dimension into
 $2D$-dimension
 after applying the two-body evolution gate $U_{zz}$, i.e.,
 the new tensors $\tilde{A}$ and $\tilde{B}$ have the bond dimension $2D$
 instead of the original bond dimension $D$.

 In order to complete updating the tensors
 for each infinitesimal time step,
 the new tensors $\tilde A$ and $\tilde B$ with the bond dimension $2D$
 in Eqs. (\ref{eq9a}) and (\ref{eq9b})
 should be reexpressed by another new tensors with the bond dimension $D$.
 This can be accomplished by using an optimal isometry $W$ that maps from $2D$- back to $D$-dimensions
 for the new tensors $\tilde A$ and $\tilde B$
 in Eqs. (\ref{eq9a}) and (\ref{eq9b}) as, respectively,
\begin{subequations}
   \label{eq4}
\begin{eqnarray}
  \sum_{l',u',r'=1}^{2D}W_{l}^{l'}W_{u}^{u'}W_{r}^{r'}\tilde{A}_{l'u'r'}^{ia}
  &=& A_{lur}^{ia}, \\
 \sum_{l',r',d'=1}^{2D}W_{l}^{l'}W_{r}^{r'}W_{d}^{d'}\tilde{B}_{l'r'd'}^{ia}
 &=& B_{lrd}^{ia}.
\end{eqnarray}
\end{subequations}
 These processes are known as the so-called renormalization
 of the updating tensors $\tilde A$ and $\tilde B$.
 Constructing the optimal isometry $W$ requires calculating
 the environment tensors of the updating tensor $\tilde{A}$ and $\tilde{B}$.
 The corner transfer matrix renormalization method \cite{Corner}
 is implemented
 to contract the environmental tensors.
 The environmental tensors are contracted with each
 other by indices of dimension $M$ (called environment dimension).
 Similar implementing processes in Ref.~\onlinecite{Czarnik1}
 have been then performed to
 get the updated tensors $A(d\beta)$
 and $B(d\beta)$ with truncating back to $D$- from $2D$-dimensions
 of the updating tensors $\tilde A$ and $\tilde B$.
%

%
\begin{figure}
\includegraphics[width =0.42\textwidth]{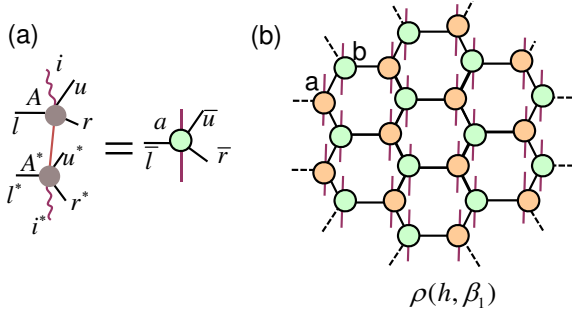}
  \caption{(color online)
     (a) Contraction of the tensor $A$ and
     the complex conjugate tensor $A^{\ast}$
     giving the reduced tensor $a$.
     (b) Tensor network representation of the density matrix
     $\rho(h,\beta)=\mathrm{Tr}_{ancillas}
     |\Psi(\beta)\rangle\langle\Psi(\beta)|$
     by tracing over the ancilla states.}
\label{fig2}
\end{figure}

\section{Thermal fidelity per lattice site}
 Once the thermal wavefunctions $|\Psi(\beta)\rangle$
 are obtained as a function of temperature
 from the finite-temperature iPEPS algorithm, the thermal density matrix
 $\rho(\beta) = \mathrm{Tr}_{ancillas}|\Psi(\beta)\rangle\langle\Psi(\beta)|$
 is obtained by taking the trace over the ancillas state of thermal wavefunctions.
 The thermal density matrix $\rho(\beta)$ can then be presented by
 the reduced tensors $a$ (denoted by orange circles) and $b$ (denoted by green circles)
 in the tensor network representation in Fig. ~\ref{fig2}(b),
 where, as is shown in Fig. \ref{fig2}(a),
 the reduced tensor $a$ is obtained by taking trace over the ancillas index of the tensor $A$
 and the complex conjugate tensor $A^*$,
 and the reduce tensor $b$ (denoted by a green circle) is calculated in the same way.
 Similar to the quantum fidelity \cite{Zhou2,Zhou08},
 the thermal fidelity can be defined
 in terms of thermal density matrices \cite{Quantum, Fidelity1, Fidelity2, Fidelity3}  as
\begin{equation}
  F(\beta_{1},\beta_{2})=\frac{\mathrm{Tr} \sqrt{\sqrt{\rho(\beta_{1})}\, \rho(\beta_{2})\, \sqrt{\rho(\beta_{1})}}}
  {\sqrt{\mathrm{Tr}\sqrt{\rho(\beta_{1})}\, \mathrm{Tr}\rho(\beta_{2})\, \mathrm{Tr}\sqrt{\rho(\beta_{1})}}}.
    \label{fidelity}
\end{equation}
 This thermal fidelity has basic properties such as
 $F(\beta,\beta) = 1$ for equal temperatures and $F(\beta_1,\beta_2) = F(\beta_2,\beta_1)$
 for exchanging the thermal states.
 Also, for relatively large lattice sites,
 the thermal fidelity can be scaled asymptotically as
 $F \sim d^{L}$, where $d$ is a scaling parameter.
 Actually,
 the scaling parameter $d$ is the averaged thermal-state fidelity per lattice site (tFLS),
 which is well defined in the thermodynamic limit,
\begin{equation}
  d(\beta_{1},\beta_{2}) \equiv \lim _{L\rightarrow\infty}F(\beta_{1},\beta_{2})^{1/L}.
    \label{dfidelity}
\end{equation}
 From the thermal fidelity, the tFLS satisfies
 (i) $d(\beta,\beta)=1$ for the normalization,
 (ii)$d(\beta_{1},\beta_{2})=d(\beta_{2},\beta_{1})$ for the exchange symmetry,
 and (iii) $0 \leq d(\beta_{1},\beta_{2})\leq 1$.
 At zero temperature $T=0$, the tFLS reduces to
 the quantum fidelity per lattice sites (FLS)  ~\cite{Zhou2,Zhou08}
 for quantum states.

    \begin{figure}
 \includegraphics[width =0.3\textwidth]{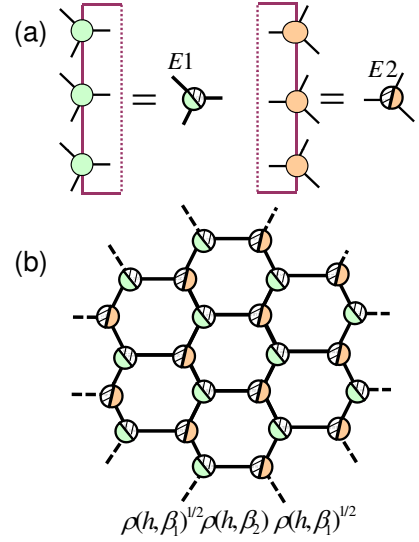}
  \caption{(color online)
      (a) Two basic cell structures for $E_1$
      and $E_2$.
      (b) Tensor network representation
      of the product $\rho(\beta_1)^{1/2} \rho(\beta_2) \rho(\beta_1)^{1/2}$.
      }
\label{fig3}
\end{figure}


 In performing the calculation of the thermal fidelity,
 for the density product, i.e., $\rho(\beta_{1})^{1/2} \rho(\beta_{2}) \rho(\beta_{1})^{1/2}$,
 the two basic cell structures can be constructed on the honeycomb lattice
 with the transfer matrices $E_1$ and $E_2$ in Fig.~\ref{fig3}(a).
 By using the two basic cell structures,
 the density product
 can be represented by contracting out the physical indices
 in the density matrix tensor in Fig.~\ref{fig3}(b).
 As a consequence, the thermal fidelity can be presented in the tensor network representation
 in Fig. \ref{fig3}(b).
 The tFLS  $d(\beta_{1},\beta_{2})$
 is equivalent to the maximum eigenvalue of the transfer matrix \cite{Zhou08}.

 Generally, in the tensor network representation of the thermal fidelity in Fig. \ref{fig3}(b),
 each bond dimension of the tensors $E_1$ and $E_2$ is $D^6$
 and then a relatively-larger environment dimension $M$ is needed for reliable calculation results.
 Consequently, calculation of the thermal fidelity in the tensor network representation in Fig. \ref{fig3}(b)
 requires a lot of computational memory space and a long calculation time.
 In our case, however, all of the system parameters of the given Hamiltonian $H$ are fixed
 in calculating the thermal fidelity.
 This fact allows us to improve the computation efficiency because
 the thermal-state fidelity can be simplified due to $\rho=e^{-\beta H}$
 as
\begin{equation}
 F(\beta_{1},\beta_{2}) = \frac{\mathrm{Tr}\, \rho(\tilde{\beta})}
 {\sqrt{\mathrm{Tr}\rho(\beta_1)} \sqrt{\mathrm{Tr}\rho(\beta_2)}},
\label{fidelity2}
\end{equation}
 where $\tilde{\beta}=(\beta_1+\beta_2)/2$.
 With the effective temperature $\tilde{\beta}=(\beta_1+\beta_2)/2$,
 the simple form of the thermal fidelity in Eq. (\ref{fidelity2})
 is represented in the tensor network representation in Fig. \ref{fig2}(b).
 In the representation, each bond dimension of the maximum tensors $a$ and $b$
 becomes $D^2$, where the tensors $a$ and $b$ correspond to the transfer matrices $E_1$ and $E_2$ in Fig. \ref{fig3}(b).
 This results in the representation dimensions of the tensors $a$ and $b$
 being much smaller than those of the tensors $E_1$ and $E_2$.
 The consequential environment dimension $M$ becomes much smaller than that in Fig. \ref{fig2} (b).
 Thus, in our study, we have used the tensor network representation in Fig. \ref{fig3}(b) of the thermal fidelity
 in Eq. (\ref{fidelity2}) with the effective temperature $\tilde{\beta}=(\beta_1+\beta_2)/2$.

 \begin{figure}
 \includegraphics[width =0.38\textwidth]{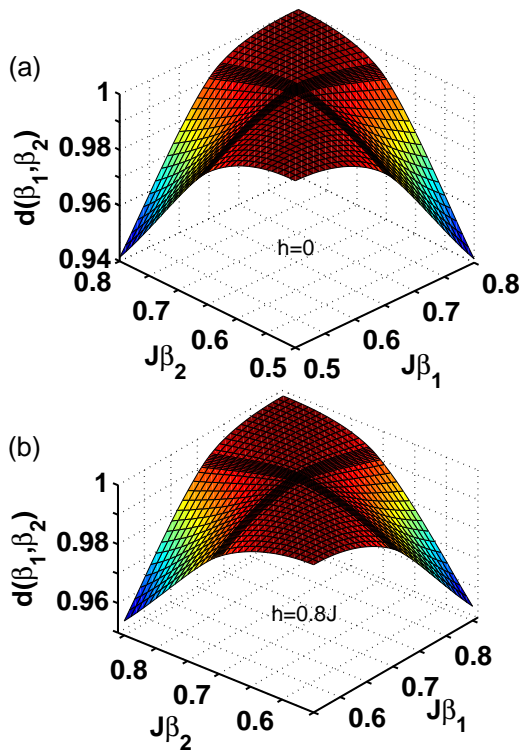}
  \caption{(color online)
   Thermal fidelity per lattice site $d(\beta_1,\beta_2)$ for transverse magnetic fields (a) $h = 0$ and (b) $h=0.8J$
   in the $\beta_1$-$\beta_2$ plane.
   A pinch point is seen in the thermal fidelity surfaces. }
\label{fig4}
\end{figure}

 \subsection{Pinch points of tFLS}
 At zero temperature, the fidelity per lattice site (FLS)
 for quantum states
 has been applied successfully in the investigations
 of quantum phase transitions because it can capture
 unstable fixed points, corresponding to phase transition points,
 along renormalization group flows
 \cite{Zhou2,Zhou08}.
 Similarly, our tFLS can capture thermal phase transition points.
 Suppose that a thermal system undergoes thermal phase transitions
 at a critical temperature $T_c$ (or $\beta_c$),
 which may imply that the thermal state of the system
 experiences a non-trivial change of its structure.
 Such a non-trivial change in the thermal state
 can be captured by the tFLS.
 Specifically, $d(\beta_1,\beta_2)$ reveals
 singular behavior when $\beta_1$ $(\beta_2)$ crosses $\beta_c$
 for a fixed $\beta_2$ $(\beta_1)$.
 At the point $(\beta_c,\beta_c)$, the singular behaviors
 can characterize a transition point, especially named as
 a pinch point $d(\beta_c,\beta_c)$ of the tFLS, which is
 the intersection of two singular lines $\beta_1 =\beta_c$
 and $\beta_2=\beta_c$ as a function of $\beta_1$ and $\beta_2$
 for continuous phase transitions.
 Then
 there are two possible ways to investigate a thermal phase transition:
 (i) detecting pinch points
 on the tFLS surface and (ii) detecting singular behavior
 of the tFLS.

 In Fig. \ref{fig4},
 we plot the tFLS surface
 $d(\beta_{1},\beta_{2})$ for (a) $h=0$ and (b) $h=0.8J$ in the $\beta_1$-$\beta_2$ plane
 for the bond dimension $D=2$ and the environment dimension $M=32$.
 In the tFLS surfaces, one can notice pinch points \cite{Zhou2,Zhou08}, which
 correspond to phase transition points, on
 intersection lines.
 From the pinch points in Fig. \ref{fig4}, we estimate the phase transition points as
 $k_B T_c =1.51745\, J$ $(J\beta_c =0.659)$ for $h=0$
 and $k_B T_c =1.40647\, J$ $(J\beta_c = 0.711)$ for $h = 0.8J$
 in the quantum transverse Ising model on the honeycomb lattice.
 We discuss the accuracy of these results in Sec. V.

    \begin{figure}
 \includegraphics[width =0.36\textwidth]{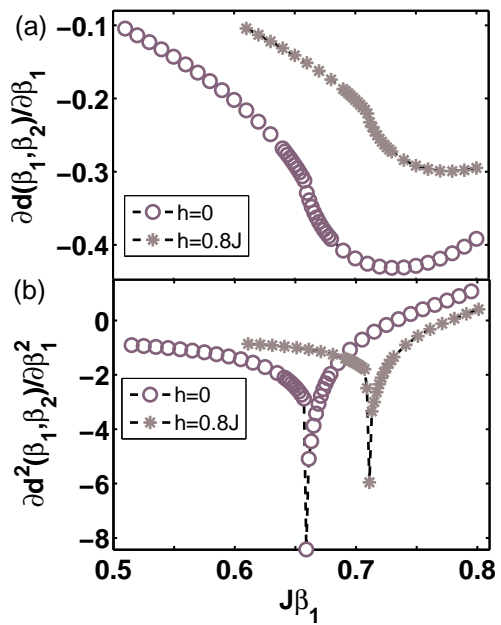}
  \caption{ (Color online)
      (a) The first partial derivative of the thermal fidelity per site $\partial d(\beta_1,\beta_2)/\partial\beta_1$
      as a function of  $\beta_1$ for transverse magnetic fields $h=0$ and $h=0.8J$
      with $J\beta_2=0.5$ and $J\beta_2=0.6$, respectively.
      (b) The second partial derivative of the thermal fidelity per site $\partial d(\beta_1,\beta_2)/\partial\beta_1$
      as a function of  $\beta_1$.
      In (b), the singular points appear
      at $J\beta_c=0.659$ and $J\beta_c=0.711$ for $h=0$ and $h=0.8J$, respectively,
      which correspond to critical points.
      }
\label{fig5}
\end{figure}

 \subsection{Singular behavior of the tFLS}
 \label{tFLS}
  As another way to determine a phase transition point from the tFLS,
  a singular behavior of the tFLS itself and its derivatives indicate
  a phase transition point.
  In order for comparison between the pinch points in determining the thermal phase points,
  let us then consider the tFLS $d(\beta_1,\beta_2)$ with a reference state $\left|\Psi (\beta_2)\right\rangle$ for a fixed value of $\beta_2$, i.e.,
  $J\beta_2 = 0.5$ for $h=0$ and $J\beta_2 = 0.6$ for $h = 0.8J$.
  In Fig. \ref{fig5}, we plot the (a) first- and (b) second-derivatives of tFLS $d(\beta,\infty)$
  as a function of $\beta$ for $h=0J$ and $h=0.8J$.
  Here, the environment truncation dimension is $M=32$
  and the step is $Jd\beta=10^{-3}$.
  The first-derivatives are shown to be continuous, i.e., to exhibit non-singular behavior.
  However, the second-derivatives exhibit singular behavior showing a discontinuity.
  The discontinuous points indicate a phase transition point, i.e., the model undergoes
  thermal phase transition across the discontinuous point of temperature.
  The discontinuous points correspond to the critical temperatures estimated as
  $k_B T_c =1.51745J$ $(J\beta_c =0.659)$ for $h=0$
  and $k_B T_c =1.40647J$ $(J\beta_c = 0.711)$ for $h = 0.8J$.
  These results indicate that both the pinch points and the singular points of the derivatives of
  the thermal fidelity give the same critical temperatures.
  Also, both the continuous fidelity surfaces in Fig. \ref{fig4} and the continuous
  behavior of the first derivative in Fig. \ref{fig5} (a) imply that the system undergoes a continuous phase transition.

\begin{figure}
 \includegraphics[width =0.36\textwidth]{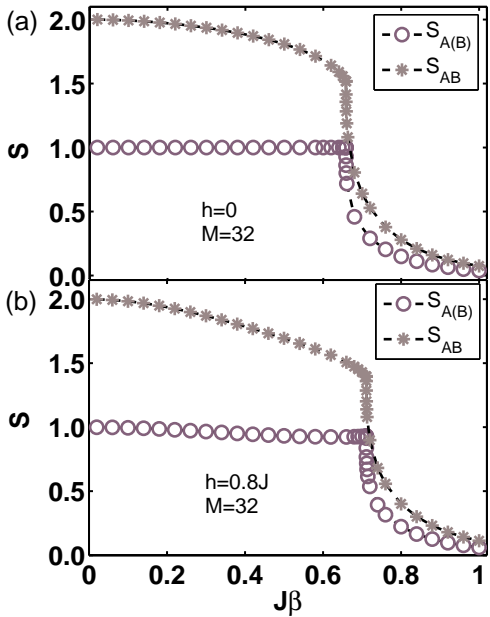}
 \caption{ (Color online)
    The von Neumann entropy $S$ as a function of inverse temperature
     $\beta$ for transverse magnetic fields (a) $h=0$ and (b) $h=0.8J$.
     The singular behavior of the von Neumann entropy is observed
      at $J\beta_c=0.6585$ and $J\beta_c=0.711$
     for transverse magnetic field $h=0$ and $h=0.8J$, respectively.
     }
     \label{SS}
     \end{figure}


\section{von Neumann entropy at finite temperature}
\label{von}

 In our tPEPS approach, we can use the thermal density matrix $\rho(h,\beta)$ in Fig. \ref{fig2}(b)
 to investigate whether finite-temperature
 phase transitions can be quantified by using the von Neumann entropy.
 We consider two types of reduced density matrices, i.e.,
 one-site reduced density matrix $\rho_{A/B} (h,\beta) = \mathrm{Tr}_{B/A \cup C}\ \rho(h,\beta)$
 and two-site reduced density matrix  $\rho_{A \cup B} (h,\beta) = \mathrm{Tr}_{C}\ \rho(h,\beta)$,
 where $C$ denotes the remainder of the system.
 The von Neumann entanglement entropy $S$ of a bipartition
 of the system is thus given in terms of the reduced density matrix
\begin{equation}
 S_j = - \mathrm{Tr}\ \rho_j (h,\beta) \log_2 \rho_j (h,\beta),
\end{equation}
where
 $\rho_j (h,\beta) = \mathrm{Tr}_{j^{\,c}} \, \rho(h,\beta)$, with $j = A, B$ or $A\cup B$,
 is the reduced density matrix obtained
 from the full density matrix
 by tracing out the degrees of
 freedom of the rest of the subsystem $j^{\, c}$.

 In Fig. \ref{SS}, we plot the von Neumann entropies as a function of the inverse temperature $\beta$
 for transverse magnetic fields (a) $h=0$ and (b) $h=0.8J$
 with the environment truncation dimension
 $M=32$ for the step $Jd\beta=10^{-4}$.
 Figure \ref{SS} shows that as temperature increases,
 both the one-site and the two-site von Neumann entropies increase
 due to the increment of thermal fluctuations
 and
 they exhibit a singular behavior.
 At the critical inverse temperatures $\beta_c$,
 the singular points correspond to the singular points of the tFLS, i.e.,
 the finite-temperature phase  transition points
 $J\beta_c=0.6585 $ and $J\beta_c=0.711 $
 for transverse magnetic field $h=0$ and $h=0.8J$, respectively.
 It is shown that the one-site and two-site von Neumann entropies
 captures the finite-temperature phase transitions
 in this model.
 Similar to the continuous behavior of quantum phase transitions \cite{Su13,Su12},
 the continuous behavior of the von Neumann entropy at the singular points
 implies that a continuous phase transition occurs
 at the transition temperatures.


 \begin{figure}
 \includegraphics[width =0.36\textwidth]{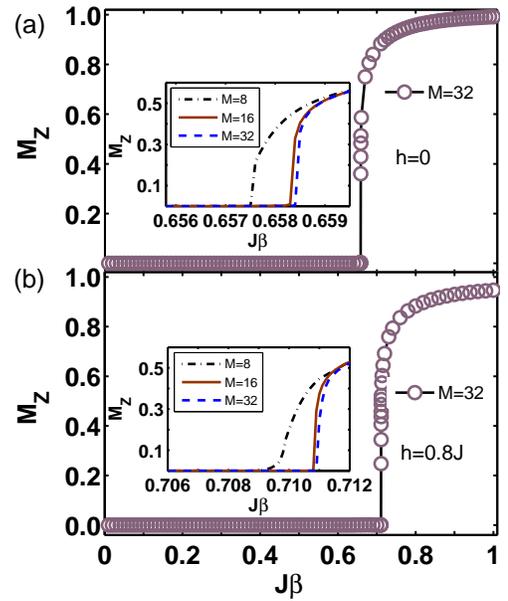}
 \caption{ (Color online)
    Magnetization $M_z$ as a function of inverse temperature
     $\beta$ for transverse magnetic fields (a) $h=0$ and (b) $h=0.8J$.
     The insets show the spontaneous magnetizations $M_Z$ plotted
     for different environment truncation dimensions $M$.
     The critical inverse temperatures are estimated as $J\beta_c=0.6585$ and $J\beta_c=0.711 J$
     for transverse magnetic field $h=0$ and $h=0.8J$, respectively.
     }
\label{fig6}
\end{figure}

\section {Transverse magnetization}
 In order to confirm the results from the tFLS and the von Neumann entropy,
 we investigate the local order parameter, defined  by
 the transverse magnetization in this section.
 In the classical limit, i.e., $\beta = 0$, for the case of $h=0$, the two site interaction
 gate $U_{zz}(\beta)$ acts on an initial state $|\Psi(0)\rangle$ and the exact state
 $|\Psi(\beta)\rangle=U_{zz}|\Psi(0)\rangle$ can be obtained.
 The bond dimension $D=2$ is then enough for an exact
 iPEPS representation of any classical state including the critical one.
 However, the calculations of expectation values require an effective approximate environment.
 Thus, in the vicinity of the critical point, a bigger environment truncation dimension $M$ is required
 to calculate expectation values of operators such as magnetizations and spin correlations~\cite{Czarnik1}.
 For the opposite limit, i.e., $\beta\rightarrow\infty$,
 which corresponds to the quantum case,
 the state of the system is in a product state configuration, where
 either every spin is in the $\left|\uparrow\right\rangle_{z}$ state or every spin is in the
 $\left|\downarrow\right\rangle_{z}$ state.
 Then the system exhibits a spontaneous
 symmetry breaking, which randomly chooses either the spin up or spin down configuration.
 According to Eqs. (\ref{eq9a}) and (\ref{eq9b}), the zero temperature ferromagnetic state
 $U_{zz}(\infty)|\Psi(0)\rangle$ is represented exactly by
 $\tilde{A}_{s_l,s_u,s_r}^{ia}\propto(-1)^{is}\delta^{ia}$
 and $\tilde{B}_{s_l,s_r,s_d}^{ia}\propto(-1)^{is}\delta^{ia}$.

 In Fig.~\ref{fig6}, we plot the magnetization $M_Z = \langle \sigma_z \rangle$
 as a function of the inverse temperature
 $\beta$ for transverse magnetic field (a) $h=0$ and (b) $h=0.8J$ with
 the environment truncation dimension
 $M=32$ for the step $Jd\beta=10^{-4}$.
 In the insets of Fig.~\ref{fig6},
 the spontaneous magnetization $M_z$ are
 plotted for different environment truncation dimension $M$.
  The spontaneous magnetizations have non-zero values
  for the inverse temperatures  $J\beta > 0.6585$ in the absence of magnetic field $h=0$
  and $J\beta > 0.711$ in the presence of magnetic field $h = 0.8J$,
  which means that the system is in the ferromagnetic phase.
  We thus obtain
  the critical inverse temperatures $J\beta_c=0.6585$ and $J\beta_c=0.711$
  for transverse magnetic fields $h=0$ and $h=0.8J$, respectively.
  These critical temperatures are consistent
  with those obtained from the tFLS in Subsec. \ref{tFLS}
  and the von Neuman entropy in Sec. \ref{von}.
  %

\begin{table*}
\renewcommand\arraystretch{2}

\caption{Critical temperature $k_BT_c (=\beta_{c})$ for values of the
 magnetic field $h$ in the honeycomb spin lattice
 with quantum Ising interaction in units of the interaction strength $J$. The fitted critical temperature
 $T^{fit}_c$ was estimated by using the phase boundary function $(k_B T_c)^2 + h^2/2 = a J^2$
 with the numerical constant $a = 2.298$. The absolute error is defined
 as $\varepsilon_{err} = \left|k_BT_c - k_BT^{fit}_c\right|$.}
\begin{tabular}{c|cccccccccc}
\hline
\hline
  $h$ & 0.0 & 0.1 & 0.2 & 0.3 & 0.4 & 0.5 & 0.6 & 0.7 & 0.8 & 0.9  \\
\hline
 $\beta_{c}$ & 0.659 & 0.659 & 0.662 & 0.665 & 0.671 & 0.678 & 0.687 &
 0.698 & 0.711 & 0.727 \\
 \hline
 $k_BT_{c}$ & 1.5175 & 1.5175 & 1.5106 & 1.5038 & 1.4903 & 1.4749 & 1.4556 &
 1.4327 & 1.4065 & 1.3755\\
 \hline
 $k_BT^{fit}_{c}$ & 1.5159 & 1.5143 & 1.5093 & 1.5010 & 1.4893 & 1.4741 & 1.4553 &
 1.4328 & 1.4064 & 1.3759\\
 \hline
 $\varepsilon_{err}$ & $1.6\times10^{-3}$ & $3.2\times10^{-3}$ & $1.3\times10^{-3}$ & $2.8\times10^{-3}$ & $1\times10^{-3}$ & $8\times10^{-4}$ & $3\times10^{-4}$ &
 $1\times10^{-4}$ & $1\times10^{-4}$ & $4\times10^{-4}$\\
 \hline
 \hline
%
 $h$ & 1.0 & 1.1 & 1.2 & 1.3 & 1.4 & 1.5 & 1.6 & 1.7 & 1.8 & 1.9  \\
\hline
 $\beta_{c}$ & 0.747 & 0.770 & 0.799 & 0.834 & 0.877 & 0.931 & 1.000 & 1.095 & 1.223 & 1.417  \\
 \hline
 $k_BT_{c}$ & 1.3387 & 1.2987 & 1.2516 & 1.199 & 1.1403 & 1.0741 &
                1.000 & 0.9132 & 0.8177 & 0.7057  \\
 \hline
 $k_BT^{fit}_{c}$ & 1.3409 & 1.3012 & 1.2562 & 1.2054 & 1.1480 & 1.0831 &
                1.009 & 0.9236 & 0.8234 & 0.7021  \\
 \hline
 $\varepsilon_{err}$ & $2.2\times10^{-3}$ & $2.5\times10^{-3}$ & $4.6\times10^{-3}$ & $6.4\times10^{-3}$ & $7.7\times10^{-3}$ & $9\times10^{-3}$ &
                $9\times10^{-3}$ & $1.04\times10^{-2}$ & $5.7\times10^{-3}$ & $3.6\times10^{-3}$  \\
 \hline
 \hline
%
\end{tabular}
\label{table}
\end{table*}


\section {Phase diagram in the presence of transverse magnetic field}
 So far we have studied the tFLS and the von Neumann entropy
 with characteristic singular behavior indicating finite-temperature
 phase transitions
 at the two magnetic field values cases $h =0$ and $h= 0.8J$
 for the quantum transverse Ising model on the honeycomb lattice.
 In this section
 we investigate the phase boundary
 in the wider parameter space.
 In determining the critical temperature and field,
 the accuracy of the iPEPS is more affected
 by the environment dimension $M$ than the bond dimension $D$.
 From our calculation, we have noticed that the practical optimized dimensions
 are the bond dimension $D=2$ and the environment dimension $M=32$ for the step $Jd\beta=10^{-3}$,
 which means that other choices for the dimensions would not change the numerical critical temperature
 within the errors of the accuracy of the iPEPS.
 As for the order parameter, the non-zero transverse magnetization also confirms
 the critical temperature and field.
%

    \begin{figure}
 \includegraphics[width =0.36\textwidth]{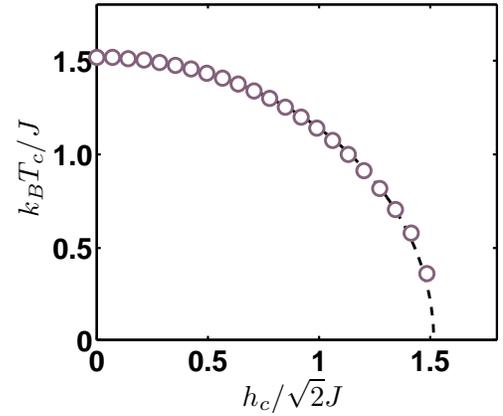}
  \caption{ Phase boundary in the temperature-magnetic field plane
            for the honeycomb lattice with quantum Ising interactions.
            The circles indicate the numerical data
            and the dashed line is the numerical fitting function
            $(k_BT_c)^2 + h_c^2/2 = a J^2$ with
            $a = 2.298$.}
 \label{fig7}
\end{figure}


 We have calculated twenty critical points including
 the case of zero-magnetic field  for the model.
 In Table \ref{table},
 we summarize the critical temperatures $k_BT_c$ and the corresponding critical
 magnetic fields $h_c$ in units of the interaction strength $J$.
 In the temperature-magnetic field plane, we plot the phase boundary in Fig. \ref{fig7}.
 As the magnetic field increases, the critical temperature becomes lower.
 Note that Fig. \ref{fig6} shows a monotonic behavior of the critical points in the temperature-magnetic field plane, which implies that the phase separation can be determined by
 a phase boundary function $f(T_c,h_c) = (k_BT_c/J)^2 + (h_c/J)^2/2$
 with
 a single numerical fitting constant $a$, i.e., $f(T_c,h_c) = a$.
 Thus the model is in the ferromagnetic phase for $f(T_c,h_c) < a$,
 with a non-magnetic phase for $f(T_c,h_c) > a$.
 A best numerical fitting is performed to give the fitting constant $a=2.298$.
 In Fig. \ref{fig7}, the dashed line is the fitted phase boundary.
 One can also estimate the critical temperature and field by using
 the fitted phase boundary $(k_BT_c)^2 + h^2_c/2 = a J^2$.
 As the magnetic field varies, the critical temperature can be obtained by the relation
 $k_BT_c =\sqrt{a J^2 - h^2/2}$.
 The critical temperatures can be estimated as, for instance, $k_BT_c = \sqrt{a} J \simeq 1.5159J$
 for zero-magnetic field $h=0$ and $k_BT_c \simeq 1.4064 J$
 for $h = 0.8$ J.
 Alternatively, as temperature varies,
 the critical field can be obtained by the relation
 $h_c =\sqrt{2 a J^2 - 2 (k_B T)^2}$.
 The critical fields can be estimated as, for instance,
 $h_c = \sqrt{2a} J \simeq 2.1438 J$ at $T=0$ and $h_c
 \simeq 2.139 J$ at $k_B T = 0.1 J$.
 For comparison with the numerical data,
 the fitted critical values are estimated with the absolute error  in the Table \ref{table}.
 Note that the numerical critical values at all points
 have the absolute errors less than around $10^{-3}$.

 Our estimated quantum critical point $h(T=0)_c$ at zero temperature
 from the phase boundary function $(k_BT_c)^2 + h^2_c/2 = a J^2$ with $a = 2.298$
 shows a good agreement with the critical value $h_c(T=0)=2.13250(4)\,J$
 estimated from the quantum Monte Carlo calculation \cite{Henk}.
 Also, our estimated critical temperature at zero-magnetic field
 is consistent with the exact value given in (\ref{exact}).
 Consequently, these results indicate that
 the phase boundary of the honeycomb lattice with the Ising interaction
 is well described by the phase boundary function
 $(k_BT_c)^2 + h^2_c/2 = a J^2$ with the single numerical fitting constant
 $a=2.298$.
 We anticipate that this curve may well be an exact result,
 with from (\ref{exact}), the constant value
 $a = 4/[\log( 2 + \sqrt{3}\, ) ]^2 = 2.3063 ...$

\section {Conclusion}
 We have investigated the phase boundary
 of the quantum transverse Ising model on the honeycomb lattice.
 To calculate the thermal groundstate at finite temperature,
 we have employed the tPEPS algorithm with ancillas.
 In order to quantify the finite-temperature phase transition,
 we have used the von Neumann entropy and
 the thermal-sate fidelity defined as the overlap measurement
 between two thermal states.
 The tensor network representation of
 the tFLS has been constructed for thermal state on the honeycomb lattice.
 The tFLS and the von Neumann entropy have been shown
 to detect successfully the phase transition points
 in the temperature-magnetic field plane.
 The phase transition points are consistent with those determined
 by the tFLS and the von Neumann entropy, which shows that the honeycomb lattice undergoes
 a continuous phase transition.
 We found that the phase boundary in the temperature-magnetic field plane
 is given by the curve $(k_BT_c)^2 + h_c^2/2 = a J^2$ with
 the single numerical fitting coefficient $a = 2.298$.
 Then for $(k_BT_c)^2 + h_c^2/2 < a J^2$,
 the model is in the ferromagnetic phase and
 for $(k_BT_c)^2 + h_c^2/2 < a J^2$, in the non-magnetic phase.
 The fitted phase boundary estimates
 the quantum critical field $h_c (T=0)= \sqrt{2a} J \simeq 2.1438 J$
 and the critical temperature $k_B T_c (h=0)= \sqrt{a} J \simeq 1.5159 J$,
 which show good agreement with
 the Monte Carlo result \cite{Henk} and the exact result (\ref{exact}).
 Similar exact curves may possibly apply for the quantum transverse Ising
 model on other planar lattices.
 Our results show that
 our thermal fidelity and von Neumann entropy for finite temperature
 can be used to capture finite-temperature phase transitions.
 Then
 the fidelity and the von Neumann entropy approaches
 can be extended to the corresponding thermal fidelity and von Neumann entropy approaches for finite temperature.

\vspace*{1cm}
\acknowledgements


 MTB gratefully acknowledges support from Chongqing University and
 the 1000 Talents Program of China.
 This work is supported in part by
 the Fundamental Research Funds for the Central Universities (Project Nos. 106112015CDJRC131215
 and 106112016CDJXY300008)
 and the National Natural Science Foundation of China (Grant Nos. 11575037, 11374379, 11674042 and 11174375).


 

\begin{thebibliography}{99}

  \bibitem{Landau}
 L. D. Landau and E. M. Lifshitz, Statistical Physics (Pergamon,
 New York, 1958).

 \bibitem{Sachdev}
  S. Sachdev, \textit{Quantum Phase Transitions} (Cambridge University,
 Cambridge, 1999).

 \bibitem{Chaikin}
 P. M. Chaikin and T. C. Lubensky, \textit{Pinciples of Condensed Matter
 Physics} (Cambridge University, Cambridge, 1995).




 \bibitem{Amico}
 L. Amico, R. Fazio, A. Osterloh and V. Vedral,
 Rev. Mod. Phys. \textbf{80}, 517 (2008).

 \bibitem{Vidal}
  G. Vidal, J. I. Latorre, E. Rico and A. Kitaev,
  Phy. Rev. Lett. \textbf{90}, 227902 (2003).

 \bibitem{Calabrese}
  P. Calabrese and J. Cardy, J. Stat. Mech. P06002 (2004).

 \bibitem{Korepin}
 V. E. Korepin, Phys. Rev. Lett. \textbf{92}, 096402 (2004).

 \bibitem{Tagliacozzo}
 L. Tagliacozzo, T. R. de Oliveira, S. Iblisdir and J. I. Latorre,
 Phys. Rev. B \textbf{78}, 024410 (2008).

 \bibitem{Pollmann}
 F. Pollmann, S. Mukerjee, A.  Turner and J. E. Moore,
 Phys. Rev. Lett. \textbf{102}, 255701 (2009).




 \bibitem{Shi}
 Q-Q. Shi, R. Or\'us, J. O. Fj{\ae}restad, and H.-Q. Zhou,
 New J. Phys. \textbf{12}, 025008 (2010).

 \bibitem{Stephan}
 J-M. St\'ephan, G. Misguich, and F. Alet,
 Phys. Rev. B \textbf{82}, 180406R (2010).

 \bibitem{Hu}
 B-Q. Hu, X-J. Liu, J-H. Liu, and H-Q. Zhou,
 New J. Phys. \textbf{13}, 093041 (2011).

 \bibitem{JHLiu}
 J-H. Liu, H-T. Wang, Q-Q. Shi, and H-Q. Zhou,
 Phys. Lett. A \textbf{376}, 2677 (2012).


  \bibitem{Affleck}
 I. Affleck and A. W. W. Ludwig,
 Phys. Rev. Lett. \textbf{67}, 161 (1991).



 \bibitem{Liu}
 X.-J. Liu, B.-Q. Hu, S. Y. Cho, H.-Q. Zhou, and Q.-Q. Shi,
 J. Korean Phys. Soc. \textbf{69}, 1212 (2016).



 \bibitem{Zanardi}
 P. Zanardi and N. Paunkovi\'{c}, Phys. Rev. E \textbf{74}, 031123 (2006);

 \bibitem{Zhou2}
 H.-Q. Zhou and J.P. Barjaktarevi\v{c},  J. Phys. A: Math. Theor. \textbf{41}, 412001 (2008).
 H.-Q. Zhou, J.-H. Zhao, and B. Li, J. Phys. A: Math. Theor. \textbf{41}, 492002 (2008).

 \bibitem{Rams}
 M. M. Rams and B. Damski, Phys. Rev. Lett. \textbf{106}, 055701 (2011).

 \bibitem{Liu2}
 J.-H. Liu, Q.-Q. Shi, J.-H. Zhao, and H.-Q. Zhou,
 J. Phys. A: Math. Theor. \textbf{44}, 495302 (2011).

 \bibitem{Gu3}
 S. Yang, S.-J. Gu, C.-P. Sun, and H.-Q. Lin, Phys. Rev. A \textbf{78}, 012304 (2008);
 S.-J. Gu, Int. J. Mod. Phys. B \textbf{24}, 4371 (2010).

 \bibitem{Gu2}
 S.-J. Gu, Chin. Phys. Lett. \textbf{26}, 026401 (2009).

 \bibitem{Xiao}
 X.-M. Lu, Z. Sun, X. Wang, and P. Zanardi, Phys. Rev. A \textbf{78}, 032309 (2008);
 X. Wang, Z. Sun, and Z. D. Wang, Phys. Rev. A \textbf{79}, 012105 (2009).


 \bibitem{Chen}
 S. Chen, L. Wang, S. J. Gu, and Y. Wang, Phys. Rev. E \textbf{76}, 061108 (2007).

 \bibitem{Mukherjee}
 V. Mukherjee and A. Dutta, Phys. Rev. B \textbf{83}, 214302 (2011).

 \bibitem{Gong}
 L. Gong and P. Tong, Phys. Rev. B \textbf{78}, 115114 (2008).

 \bibitem{Wang}
 H.-L. Wang, J.-H. Zhao, B. Li, and H.-Q. Zhou, J. Stat. Mech.,
 L10001 (2011).

  \bibitem{Wang13}
 H. T. Wang, B. Li, and S. Y. Cho,
 Phys. Rev. B \textbf{87}, 054402 (2013).

 \bibitem{Su13}
 Y.H. Su, B.-Q. Hu, S.-H. Li and S. Y. Cho,
 Phys. Rev. E \textbf{88}, 032110 (2013).


 \bibitem{Dai}
 Y.-W. Dai, S. Y. Cho, M. T. Batchelor, and H.-Q. Zhou,
 Phys. Rev. E \textbf{89}, 062142 (2014)


\bibitem{Orus16}
 P. Schmoll and R. Or\'us,
 arXiv:1605.04315.

\bibitem{JunpengCao}
 J. Cao, X. Cui, Z. Qi, W. Lu, Q. Niu, and Y. Wang,
 Phys. Rev. B \textbf{75}, 172401 (2007).

\bibitem{Chamon}
 C. Castelnovo and C. Chamon,
 Phys. Rev. B \textbf{76}, 184442 (2007).

\bibitem{Popkov}
 V. Popkov and M. Salerno,
 Europhys. Lett. \textbf{84}, 30007(2008).


\bibitem{YangZhao}
 Y. Zhao, W. Li, B. Xi, Z. Zhang, X. Yan, S.-J. Ran, T. Liu, and G. Su,
 Phys. Rev. E \textbf{87}, 032151 (2013).

\bibitem{Porter}
 W. J. Porter and J. E. Drut, Phys. Rev. B \textbf{94}, 165112 (2016).


\bibitem{iPEPS} J. Jordan, R. Or\'us, G. Vidal, F. Verstraete and J. I. Cirac,
             Phys. Rev. Lett. \textbf{101}, 250602 (2008);
             H. C. Jiang, Z. Y. Wang and T. Xiang, Phys. Rev. Lett. \textbf{101}, 090603 (2008);
             R. Or\'us and G. Vidal, Phys. Rev. B \textbf{78}, 155117 (2008).




\bibitem{Czarnik1} P. Czarnik, L. Cincio, and J. Dziarmaga,
    Phys. Rev. B, \textbf{86}, 245101 (2012);
    P. Czarnik and J. Dziarmaga, \textit{ibid.} \textbf{90}, 035144 (2014);
    P. Czarnik and J. Dziarmaga, \textit{ibid.} \textbf{92}, 035120 (2015).

\bibitem{Czarnik4} P. Czarnik and J. Dziarmaga, Phys. Rev. B, \textbf{92}, 035152 (2015);
    P. Czarnik, J. Dziarmaga, and A. M. Ole\'s, \textit{ibid.} \textbf{93}, 184410 (2016).
    P. Czarnik, M. M. Rams, and J. Dziarmaga, \textit{ibid.} \textbf{94}, 235142 (2016).


 \bibitem{Henk}
 H. W. J. Bl\"ote and Y. Deng, Phys. Rev. E. \textbf{66}, 066110 (2002).



 \bibitem{Baxter}
 R. J. Baxter, \textit{Exactly Solved Models in Statistical Mechanics}
 (Academic Press, London, 1982), Chap. 6.

 \bibitem{Houtappel}
  R. M. F. Houtappel, Physica \textbf{16}, 425 (1950).



 \bibitem{iTEBD} G. Vidal, Phys. Rev. Lett. \textbf{98}, 070201 (2007).

 \bibitem{Suzuki} M. Suzuki, Phys. Lett. A \textbf{146}, 319 (1990).

 \bibitem{Corner} R. Or\'us and G. Vidal, Phys. Rev. B \textbf{80}, 094403 (2009).





\bibitem{Zhou08} H.-Q. Zhou, R. Or\'{u}s and G. Vidal,
                 Phys. Rev. Lett. \textbf{100}, 080602 (2008).




\bibitem{Quantum}
 M. A. Nielsen and I. L. Chuang,
 \textit{Quantum computation and quantum information}
 (Cambridge University Press, Cambridge, 2000)
                  .
\bibitem{Fidelity1}
 H. T. Quan and F. M. Cucchietti,
 Phys. Rev. B \textbf{79}, 031101 (2009).

\bibitem{Fidelity2}
 P. Zanardi, H. T.Quan, Xiaoguang Wang, and C. P. Sun,
 Phys. Rev. A \textbf{75}, 032109 (2007).

\bibitem{Fidelity3}
 J. Sirker, Phys. Rev. Lett. \textbf{105}, 117203 (2010).



 \bibitem{Su12}
  Y. H. Su, S. Y. Cho, B. Li, H. L. Wang, and H.-Q. Zhou,
  J. Phys. Soc. Jpn. \textbf{81}, 074003 (2012).


%




 \end{thebibliography}
\end{document}